\documentclass[pra,floatfix,twocolumn]{revtex4}
\usepackage{pstricks,graphicx,amsmath,bbm,mathrsfs,amssymb,times,psfrag,pifont}
\newcommand{\ket}[1]{\vert #1 \rangle} 
\newcommand{\dket}[1]{\vert #1 \rangle\rangle} 
\newcommand{\braket}[2]{\langle #1 \vert #2 \rangle}
\newcommand{\bmsigma}{\boldsymbol \sigma}

\begin{document}
\title{Entanglement induced transparency and applications}
\author{Stefano Olivares}
\email{stefano.olivares@mi.infn.it}
\affiliation{CNISM UdR Milano Universit\`a, I-20133 Milano, Italy}
\affiliation{Dipartimento di Fisica, Universit\`a degli Studi di Milano, I-20133 Milano, Italy}
\author{Matteo G. A. Paris}
\email{matteo.paris@fisica.unimi.it}
\affiliation{Dipartimento di Fisica, Universit\`a degli Studi di Milano, I-20133 Milano, Italy}
\affiliation{CNISM UdR Milano Universit\`a, I-20133 Milano, Italy}
\affiliation{ISI Foundation, I-10133 Torino, Italy}
\date{\today}
\begin{abstract}
We point out a symmetry exhibited by pairs of entangled states and
discuss its possible applications in quantum information. More specifically,
we consider quadripartite systems prepared in bipartite product states
of the form $|\Psi\rangle = |\psi\rangle_{12} \otimes |\psi\rangle_{34}$
and let the uncorrelated subsystems $14$ and $23$ interact by a given 
unitary $U_{14}\otimes U_{23}$: we show that the entanglement between the 
noninteracting subsystems $12$ and $34$ may induce transparency, i.e. 
makes $|\Psi\rangle$ an eigenstate of the unitary. We investigate the 
occurrence of this phenomenon both in continuous variable and qubit 
systems, and discuss its possible applications to bath engineering, 
double swapping and remote inversion. 
\end{abstract}
\maketitle
\section{Introduction}\label{s:intro}
Entanglement is a relevant resource for quantum information processing
and considerable efforts in this field have been devoted to investigate
its generation, characterization, manipulation and storing
\cite{eres,edet,eman}.  It is a general fact that the interaction with
external systems may lead to entanglement degradation and, in turn, much
attention has been payed to design and implement schemes suitable to
preserve and restore entanglement \cite{BB1,BB2}. Among the other
approaches, the analysis of symmetries is a powerful tool to investigate
the separability problem and the dynamics of entanglement, as well as to
individuate  systems of interest for quantum information processing, as
for example decoherence-free subspaces \cite{DFS97} and cluster states
\cite{cl1}.  In this paper we follow the above intuition and exploit a
specific symmetry exhibited by pairs of entangled states for
applications to bath  engineering, double swapping and remote
inversion. 
\par
The basic idea is to consider quadripartite systems prepared in
bipartite product states of the form $|\Psi\rangle = |\psi\rangle_{12}
\otimes |\psi\rangle_{34}$ and let the uncorrelated systems $14$ and
$23$ interact by unitaries of the form $U_{14}\otimes U_{23}$. As we
will see, there are conditions leading to {\em entanglement induced
transparency},  namely conditions in which entanglement of the
systems $12$ and $34$  preserve the initial state and its properties
during the evolution.  In other words, the overall state $|\Psi\rangle$
becomes an eigenstate of $U_{14} \otimes U_{23}$.  In the following we
investigate the occurrence of this phenomenon both in continuous variable
and qubit systems, and discuss possible applications in quantum
information processing.
\par
The paper is structured as follows. In section \ref{s:phantom} we
describe entanglement induced transparency in continuous variable
systems and investigate in details the peculiar role of twin beams, 
{\em i.e} maximally entangled states at fixed energy. We also give the 
phase-space analysis of the effect to discuss its robustness to 
pertubations and the evolution of entanglement of formation. In 
Section \ref{s:BE} we address application of entanglement 
induced transparency to bath engineering. In Section \ref{s:RI}, 
we consider the discrete variable counterpart of entanglement induced 
transparency and describe its application to the remote inversion of 
an operation acting on two qubits belonging to two different entangled 
states. Section \ref{s:remarks} closes the paper with some concluding 
remarks.
\section{Entanglement induced transparency in continuous variable
systems}\label{s:phantom}
Let us first consider continuous variable systems to illustrate
the effect of entanglement induced transparency (EntIT). In particular 
we show the crucial role played by twin-beam (TWB) entanglement.
We address a system composed by four modes, with 
field operators $a_k$, $k=1,..,4$ and consider quadripartite (pure) state 
of the form
\begin{equation}
\ket{\psi_{\rm in}} = \sum_{n,m}\psi_{n,m} \ket{n}_1 \otimes \ket{m}_2
\otimes \sum_{h,k} \omega_{h,k} \ket{h}_3 \otimes \ket{k}_4\:.
\end{equation}
Then we make the modes 14 and 23 interact through bilinear Hamiltonians,
as those describing the interaction of modes in a beam splitter. 
We denote by $U_{14}(\phi)$ and $U_{23}(\varphi)$ the corresponding 
unitary operations, which lead to the following Heisenberg evolution  
\begin{subequations}
\label{genBS}
\begin{align}
U_{hk}^{\dag}(\alpha) a_k U_{hk}(\alpha) &= a_h\cos\alpha+a_k\sin\alpha \\
U_{hk}^{\dag}(\alpha) a_h U_{hk}(\alpha) &= - a_h\sin\alpha+a_k\cos\alpha
\end{align}
\end{subequations}
for the field modes. 
This corresponds 
to have beam splitters with transmissivities $T_{14} = \cos^2 \phi$ 
and $T_{23} = \cos^2 \varphi$. Upon expanding Fock number states 
$\ket{n}_k = (n!)^{-1/2} (a_k^\dag)^n\ket{0}$ and using mode
transformation  (\ref{genBS}) one obtains:
\begin{align}
\ket{\psi_{\rm out}}&= 
U_{14}(\phi)\otimes U_{23}(\varphi)\ket{\psi_{\rm in}}\\
&= 
\sum_{n,m,h,k} \frac{\psi_{n,m}\,\omega_{h,k}}{\sqrt{n!\, m!\, h!\, k!}}
\nonumber\\
\times&
\sum_{s=0}^{n} \sum_{t=0}^{m} \sum_{r=0}^{h} \sum_{u=0}^{k} (-1)^{r+u}
{n\choose s}  {m\choose t} {h\choose r}  {k\choose u}
\nonumber\\
\times&
\sqrt{(n-s+u)!\, (m-t+r)!\, (h-r+t)!\, (k-u+s)!} \nonumber\\
\times&
(\cos\phi)^{n+k-s-u}
(\sin\phi)^{s+u}
(\cos\varphi)^{m+h-t-r}
(\sin\varphi)^{t+r}
\nonumber\\
\times&
\ket{n-s+u}_1
\ket{m-t+r}_2
\ket{h-r+t}_3
\ket{k-u+s}_4
\label{cumber}
\end{align}
where, for the sake of simplicity, we omitted the tensor product symbol
in the last line.
The conditions for transparency $\ket{\psi_{\rm out}}=\ket{\psi_{\rm in}}$,
can be obtained by imposing
$\braket{nmhk}{\psi_{\rm in}}=\braket{nmhk}{\psi_{\rm out}}$ for all the 
basis elements $\ket{nmhk}\equiv 
|n\rangle_1 |m\rangle_2 |h\rangle_3 |k\rangle_4$). It turns out that the
whole set of equations is subsumed by a single condition 
\begin{align}
\psi_{1,1}\,\omega_{1,1} =&
\cos(2\phi)\cos(2\varphi) \psi_{1,1}\,\omega_{1,1}\nonumber\\
&+\frac{\cos(2\phi)\sin(2\varphi)}{\sqrt{2}}
(\psi_{1,0}\,\omega_{2,1} - \psi_{1,2}\,\omega_{0,1})\nonumber\\
&+\frac{\sin(2\phi)\cos(2\varphi)}{\sqrt{2}}
(\psi_{0,1}\,\omega_{1,2} - \psi_{2,1}\,\omega_{1,0})\nonumber\\
&+\frac{\sin(2\phi)\sin(2\varphi)}{2}\nonumber\\
\times&
(\psi_{2,2}\,\omega_{0,0} - \psi_{0,2}\,\omega_{0,2} -
\psi_{2,0}\,\omega_{2,0} + \psi_{0,0}\,\omega_{2,2}).
\label{cond1111}
\end{align}
If $\psi_{n,m}$ and $\omega_{h,k}$ are generic, then
the only solution of (\ref{cond1111})
is $\phi=\varphi=0$, i.e., the BSs transmissivity should be equal to 1
(trivial solution).
For completeness, we note that also $\phi=\varphi=\pi/2$ is a solution, but,
in this case, the two mode are just exchanged. In particular, these 
results can be applied to separable states,
in which $\psi_{n,m}$ and
$\omega_{h,k}$ can be written as product of two terms respectively. 
In other words, classical correlations are not enough to obtain the 
transparency. 
\par Let us now consider the case of photon-number entangled 
states, namely states with  $\psi_{n,m} \propto \delta_{nm}\, \psi_n$
and $\omega_{h,k} = \delta_{kh}\, \psi_k$, $\delta_{nm}$ being the
Kronecker delta. TWBs belong to this class, as well as the so-called 
pair-coherent states \cite{aga86,aga05}, which find
applications in quantum communication \cite{us1,us2}. In this case 
Eq.~(\ref{cond1111}) reduces to:
\begin{align}
\psi_1\, \omega_1 =& \cos(2\phi)\cos(2\varphi)\,\psi_1\, \omega_1 \nonumber\\
&+ 2 \cos(\phi)\sin(\phi)\cos(\varphi)\sin(\varphi)\nonumber\\
&\times(\psi_2\,\omega_0 + \psi_0\, \omega_2)\:,
\label{cond:PNES}
\end{align}
which, in general, admits only the trivial solution $\phi=\varphi=0$.
However, if one choose $\psi_n = \omega_n = \lambda^n$, corresponding to
a pair of TWB states, Eq.~(\ref{cond:PNES}) simplifies to 
$\cos[2(\phi-\varphi)]=1$ i.e., one has the transparency effect by 
choosing $\phi=\varphi$.  
We can conclude that only TWB states may give rise to the EntIT due to
their peculiar analytical expression and, in turn, the transparency is
induced by the TWB entanglement. It is worth to note that when EntIT
occurs, then the quadripartite input state is an eigenstate of the
unitary transformation $U_{14}(\phi)\otimes U_{23}(\phi)$.
Notice that EntIT takes place for {\em any} value of the beam splitters
transmissivity provided they are equal.
\par
\begin{figure}[h]
\includegraphics[width=0.3\textwidth]{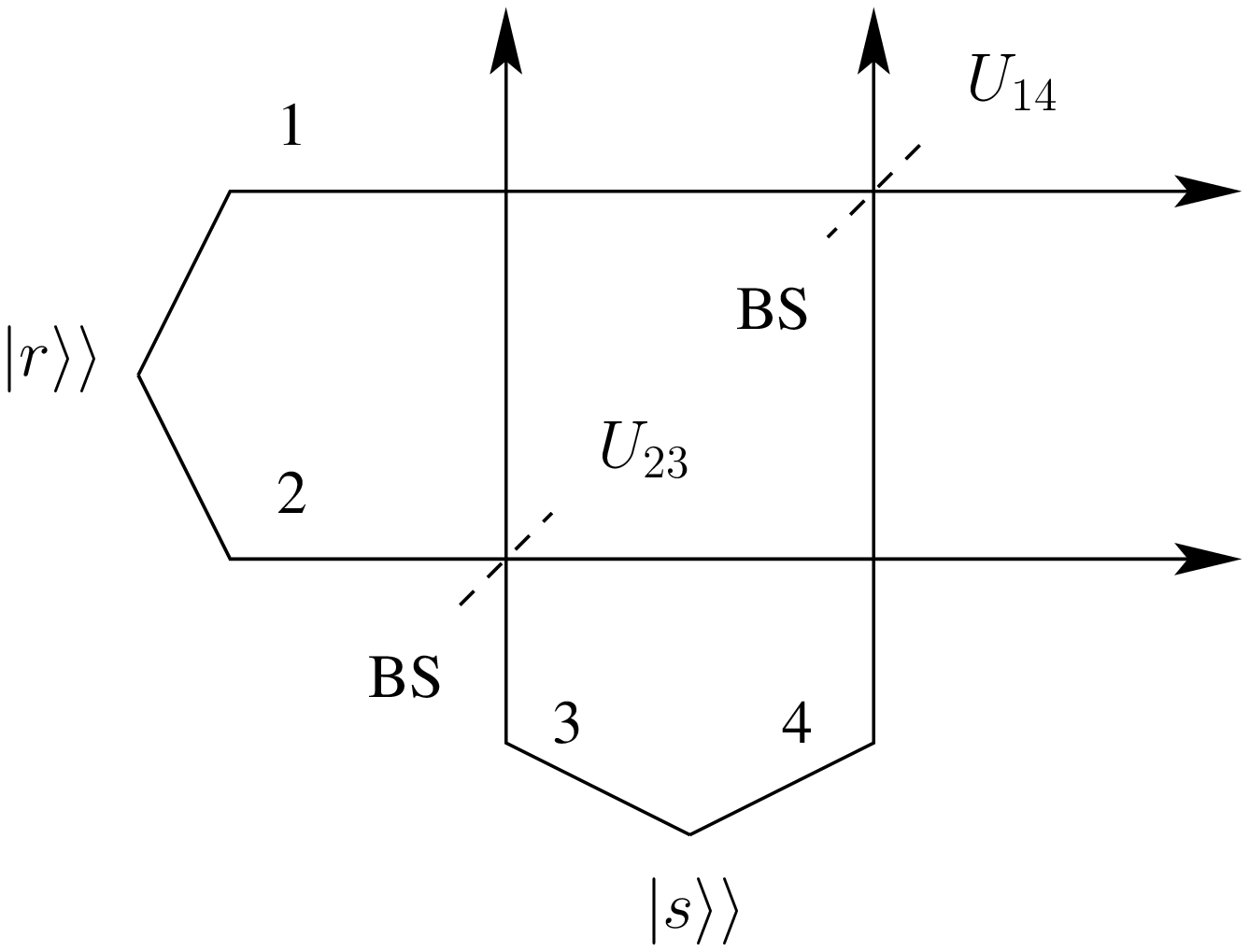}
\vspace{-0.2cm}
\caption{\label{f:scheme} Entanglement induced transparency/swapping 
when the pair of TWB, $\dket{r}$ and $\dket{s}$, are mixed at two balanced beam 
splitters.}
\end{figure}

In the following, we focus the attention on the case of a pair of 
TWB, $\dket{r}$ and $\dket{s}$ for 
modes 12 and 34 and mix modes 14 and 23 of state in two {\em balanced} 
BSs (Fig.~\ref{f:scheme}). The initial state
$\ket{\psi_{\rm in}}$ is given by two TWB states with (real)
parameters $r$ and $s$, namely:
\begin{equation}\label{in}
\ket{\psi_{\rm in}} = 
S_{12}(r)\otimes S_{34}(s) 
\ket{0}
= \dket{r}\otimes\dket{s},
\end{equation}
where
$S_{hk}(\zeta) = \exp(\zeta a_h^{\dag}a_k^{\dag} - h.c.)$ is the
two-mode squeezing operator acting onto modes $h$ and $k$, respectively,
and $a_l$ is the annihilation operator of mode $l$.
 \\
Thanks to the transformations (\ref{genBS}), the state emerging from
the BSs can be written as:
\begin{align}
\ket{\psi_{\rm out}} =& 
U_{14}\otimes U_{23}
\ket{\psi_{\rm in}} \\
=&\exp\Bigg\{\frac12\Big[
(r+s)(a_1^{\dag}a_2^{\dag}+a_3^{\dag}a_4^{\dag})\nonumber\\
& + (r-s)(a_1^{\dag}a_3^{\dag}+a_2^{\dag}a_4^{\dag}) - h.c.
\Big]\Bigg\} \ket{0}.\label{out}
\end{align}
Form Eq.~(\ref{out}) we see that:
\begin{align}
&\mbox{if } s=r \quad &\Rightarrow \quad
&\ket{\psi_{\rm out}} = \ket{\psi_{\rm in}}; \label{phantom}\\
&\mbox{if } s=-r \quad &\Rightarrow \quad
&\ket{\psi_{\rm out}} = S_{13}(r) S_{24}(r)\ket{0}. \label{swap}
\end{align}
In other words, in the case (\ref{phantom}) we have
the EntIT effect, i.e.,
$\ket{\psi_{\rm in}} \to \ket{\psi_{\rm in}}$; in the case
(\ref{swap}) we obtain
$S_{12}(r)\otimes S_{34}(r)\ket{0} \to S_{13}(r)\otimes S_{24}(r)\ket{0}$,
i.e., entanglement is ``swapped'' from mode 12 and 34 to modes
13 and 24 (note that modes 1 and 3, as well as modes 2 and 4, did not
directly interact each other). In the last case no measurement is
required to obtain (double) entanglement swapping, even if to this
aim one should have two {\em identical} states as inputs 
(double swapping).
More generally, if the BSs have transmissivity
$T_{14} = \cos^2 \phi$ and $T_{23} = \cos^2 \varphi$, then the outgoing
state reads:
\begin{align}
\ket{\psi_{\rm out}}=&
\exp\big[
(r \cos\phi \cos\varphi + s \sin\phi\sin\varphi)a_1^{\dag}a_2^{\dag}\nonumber\\
&+(r \sin\phi \sin\varphi + s \cos\phi\cos\varphi)a_3^{\dag}a_4^{\dag}\nonumber\\
&+(r \cos\phi \sin\varphi - s \sin\phi\cos\varphi)a_1^{\dag}a_3^{\dag}\nonumber\\
&+(r \sin\phi \cos\varphi - s \cos\phi\sin\varphi)a_2^{\dag}a_4^{\dag}-h.c.
\big]\ket{0},\label{out:2}
\end{align}
which reduces to (\ref{out}) if $\phi=\varphi=\pi/4$. Investigating
Eq.~(\ref{out:2}), we see that if $\varphi=\phi$ we have:
\begin{align}
\ket{\psi_{\rm out}}=&
\exp\big[
(r \cos^2\phi + s \sin^2\phi)a_1^{\dag}a_2^{\dag}\nonumber\\
&+(r \sin^2\phi + s \cos^2\phi)a_3^{\dag}a_4^{\dag}\nonumber\\
&+(r-s) \cos\phi \sin\phi (a_1^{\dag}a_3^{\dag}+a_2^{\dag}a_4^{\dag})-h.c.
\big]\ket{0},\label{out:3}
\end{align}
and one can always obtain the EntIT for $r=s$, whereas there is
no way to obtain {\em perfect} swapping.
\subsection{Phase-space analysis and robustness of EnIT}
\label{s:PSanalysis}
\begin{figure}
\includegraphics[width=0.3\textwidth]{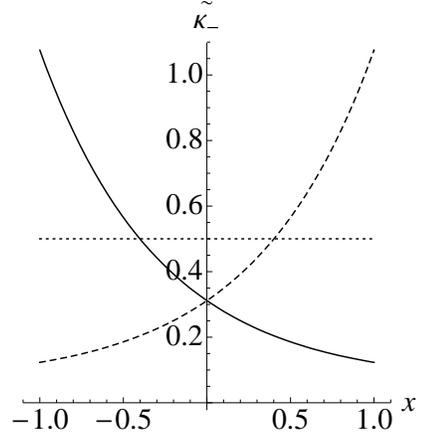}\\
\caption{\label{f:km} Plot of the minimum symplectic eigenvalue
$\tilde\kappa_{-}$ of (partially transposed) density matrices 
$\varrho^{(12)}$ (solid line) and $\varrho^{(13)}$ (dashed) 
as functions of $x=s/r$.
We chose  two balanced BSs, i.e.,
$\phi = \varphi = \pi/4$, and put $r=0.7$. See the text for details.
The dotted line refers to the separability threshold
$\tilde\kappa_{-} = 1/2$: if $\tilde\kappa_{-} \ge 1/2$ the state is
separable.}
\end{figure}

In this section we characterize the state (\ref{out:3}) in the
phase-space in order to address the distribution of two-mode 
entanglement among all the possible partitions and the robustness
of the EntIT effect.
In order to simplify the formalism, we will use the following notation 
for the input/output
modes $h,k,\ldots$: when we write ``$\square_{hk\ldots}$'', we refer
to input modes, whereas writing ``$\square^{(hk\ldots)}$'' output
modes are considered.
Since the involved states are Gaussian and the evolution preserves this
character, in order to characterize the output states we consider the
evolution of the covariance matrix (CM) \cite{gsb}. The CM associated with
the two-mode squeezed vacuum state $\dket{\psi_{hk}(r)} = S_{hk}(r)\ket{0}$
of modes $h$ and $k$ reads:
\begin{equation}
{\boldsymbol \Sigma}_{hk}(r) = \frac12\left(
\begin{array}{c|c}
\cosh 2r\, {\mathbbm 1}_2 & \sinh 2r\, {\bmsigma}_3 \\
\hline
-\sinh 2r\, {\bmsigma}_3 & \cosh 2r\, {\mathbbm 1}_2
\end{array}
\right),
\end{equation}
where $ {\mathbbm 1}_2$ is the $2\times 2$ identity matrix and
${\bmsigma}_3 = \mbox{Diag}(1,-1)$ is the Pauli matrix.
Thus, the four-mode covariance matrix of state (\ref{in}) is given by:
\begin{equation}
{\boldsymbol \Sigma}_{1234} =
{\boldsymbol \Sigma}_{12}(r)\oplus{\boldsymbol \Sigma}_{34}(s) =
\left(
\begin{array}{c|c}
{\boldsymbol \Sigma}_{12}(r) & {\boldsymbol 0} \\
\hline
{\boldsymbol 0} & {\boldsymbol \Sigma}_{34}(s)
\end{array}
\right).
\end{equation}
The CM after the evolution through the BSs can be obtained as follows:
\begin{equation}
{\boldsymbol \Sigma}^{(1234)}(\phi,\varphi) =
{\boldsymbol S}^{T}(\phi,\varphi) {\boldsymbol \Sigma}_{1234}
{\boldsymbol S}(\phi,\varphi)
\end{equation}
where:
\begin{align}
{\boldsymbol S}(\phi,\varphi) =
\left(
\begin{array}{c|c|c|c}
\cos\phi\,{\mathbbm 1}_2 & {\boldsymbol 0} & 
{\boldsymbol 0} & \sin\phi\,{\mathbbm 1}_2 \\
\hline
{\boldsymbol 0} &\cos\varphi\,{\mathbbm 1}_2 &
\sin\varphi\,{\mathbbm 1}_2 & {\boldsymbol 0} \\
\hline
{\boldsymbol 0} &-\sin\varphi\,{\mathbbm 1}_2 &
\cos\varphi\,{\mathbbm 1}_2 & {\boldsymbol 0} \\
\hline
-\sin\phi\,{\mathbbm 1}_2 & {\boldsymbol 0} & 
{\boldsymbol 0} & \cos\phi\,{\mathbbm 1}_2 \\
\end{array}
\right),
\end{align}
is the symplectic transformation associated with mode transformation
(\ref{genBS}).
Now, if we use the following $2\times 2$ block matrix decomposition
of a $8\times 8$ matrix:
\begin{equation}
{\boldsymbol A} = 
\left(
\begin{array}{c|c|c|c}
{\boldsymbol A}_{11} & {\boldsymbol A}_{12} &
{\boldsymbol A}_{13} & {\boldsymbol A}_{14} \\
\hline
{\boldsymbol A}_{21} & {\boldsymbol A}_{22} &
{\boldsymbol A}_{23} & {\boldsymbol A}_{24} \\
\hline
{\boldsymbol A}_{31} & {\boldsymbol A}_{32} &
{\boldsymbol A}_{33} & {\boldsymbol A}_{34} \\
\hline
{\boldsymbol A}_{41} & {\boldsymbol A}_{42} &
{\boldsymbol A}_{43} & {\boldsymbol A}_{44}
\end{array}\right),
\end{equation}
and introduce the notation:
\begin{equation}
[[{\boldsymbol A}]]_{hk} \equiv 
\left(
\begin{array}{c|c}
{\boldsymbol A}_{hh} & {\boldsymbol A}_{hk} \\
\hline
{\boldsymbol A}_{kh} & {\boldsymbol A}_{kk} \\
\end{array}
\right),
\end{equation}
then the CM ${\boldsymbol \Sigma^{(hk)}}$ associated with the reduced state
$\varrho^{(hk)} = \mbox{Tr}_{n,m}[\varrho^{(1234)}]$, with $n\ne m\ne h\ne k$,
$\varrho_{1234}$ being the density matrix of the state (\ref{out:2}):
\begin{equation}\label{reduced}
{\boldsymbol \Sigma^{(hk)}} = [[{\boldsymbol \Sigma^{(1234)}}]]_{hk}.
\end{equation}
Starting from ${\boldsymbol \Sigma^{(hk)}}$ one can easily evaluate
the purity, the separability and the entanglement of formation
of the reduced states $\varrho^{(hk)}$. As an example, we plot
in Fig.~\ref{f:km} the minimum symplectic eigenvalue $\tilde\kappa_{-}$
of the partially transposed of the reduced density matrices $\varrho^{(12)}$,
$\varrho^{(13)}$  as a function of $x\in [-1,1]$, where we fixed $r$ and 
put $x=s/r$. For the reduced density matrices $\varrho^{(34)}$ and 
$\varrho^{(24)}$ one has the same results. 
Recalling that a bipartite
Gaussian state is separable iff $\tilde\kappa_{-} \ge 1/2$, from
Fig.~\ref{f:km} we can see the swapping of entanglement; notice that there is
an interval of values of $x$ for which all the four partitions are not
separable.
In the present case, due to the symmetry of the reduced states, their
entanglement of formation is given by \cite{EOFM}
\begin{equation}
E_f=\left( \chi + \frac12 \right) \ln \left( \chi + \frac12 \right) -
\left( \chi - \frac12 \right) \ln \left( \chi - \frac12 \right),
\end{equation}
where $\chi = (\tilde\kappa_{-}^2+1/4)/(2 \tilde\kappa_{-})$.
We have seen that EntIT is achieved requiring $s=r$ and $\varphi=\phi$.
In order to evaluate the robustness of the effect, we address the fidelity
between $\varrho_{12}$ and $\varrho^{(12)}$, the input and the output state
of modes 1 and 2, respectively. Since they are both Gaussian states, the
fidelity is given by:
\begin{equation}
F=\left\{\hbox{Det}\left[\boldsymbol{\Sigma}_{12}+
\boldsymbol{\Sigma}^{(12)}\right]\right\}^{-1/2}.
\end{equation}
The analytic expression of $F$ is quite cumbersome; however, we report
two relevant series expansions with respect to the TWB 
parameters ($r$ and $s$) and the BSs transmissivities
($T_{14} = \cos^2 \phi$ and $T_{23} = \cos^2 \varphi$).
The first expansion concerns the TWB parameters in the case
$\varphi = \phi$ and reads:
\begin{equation}\label{sq:exp}
F \approx 1 - \frac12 \left[ 3 + \sin^2 \phi \, \cos(2\phi) \right] (s-r)^2;
\end{equation}
the second one addresses the BSs transmissivities:
\begin{equation}\label{BS:exp}
F \approx f + f^2\, \sin(2\phi)\, \cos^2(\phi)\, \sinh^2(r-s)\, (\phi-\varphi), 
\end{equation}
where $f\equiv f(r,s,\phi)$ is given by:
\begin{equation}
f = \frac{1}{1+[1 + \cos^2(\phi)] \sin^2(\phi) \sinh^2(r-s)}.
\end{equation}
Eq.~(\ref{sq:exp}) shows the robustness of the effect with respect to
fluctuations of $s$, whereas, from Eq.~(\ref{BS:exp}), we conclude that
the fluctuations of BSs transmissivities may be quite relevant.
\section{Application to bath engineering}\label{s:BE}
It is well known that correlated noise may enhance the capacity of a
quantum channel \cite{CM02,KB04}. Moreover, in the last years the
bosonic quantum channels with memory have attracted a growing interest
\cite{DK05,NJC05,CL09}. These channels are characterized by Gaussian
distributed-thermal noise and correlations between the environmental
modes. Even if only classical correlations have been addressed so far,
it has been demonstrated that uncorrelated phase-sensitive
environments, i.e, uncorrelated noisy channels with squeezed
fluctuations can been addressed for preservation of macroscopic
quantum coherence \cite{TABK88,PT94,NL98} or for improving
teleportation of squeezed states \cite{SO03}.  A new insight to the
properties of this kind of channels, when also nonclassical
correlations are present, may be given by applying our analysis to a
simple case of bath engineering, where entangled bath oscillators let
properties such as entanglement and purity of an input state survive
longer than uncorrelated ones do.
\begin{figure}
\includegraphics[width=0.4\textwidth]{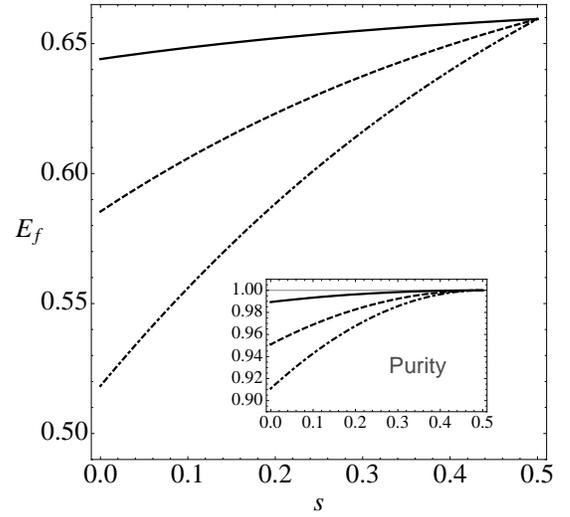}
\vspace{-0.3cm}
\caption{\label{f:EfP} Entanglement of formation,
$E_f$, of the state $\varrho^{(12)}$ a function of the TWB
parameter $s$ used to engineer the coupling (see Fig.~\ref{f:scheme})
for $r=0.5$ and different values of loss parameter $\Gamma$, from
top to bottom: $\Gamma=0.01$ (solid), $0.05$ (dashed) and $0.1$ (dot-dashed).
The inset shows the corresponding value of the purity as a function of $s$.}
\end{figure}
Let us now assume that the BSs in Fig.~\ref{f:scheme} describe
linear losses (with the vacuum as input for both the ports 3 and 4,
i.e., $s=0$). In this case, it is useful to define
$\phi = \arccos\sqrt{1-\Gamma}$, where $T=\cos^2\phi$ is the
BS transmissivity (we assume that both the BS have the same
transmissivity, i.e., $\varphi=\phi$):
if $\Gamma=0$ there aren't losses, is $\Gamma=1$ the
state is completely lost. We will refer to $\Gamma$ as loss parameter.
As a matter of fact, losses degrade the properties of the outgoing
state $\varrho^{(12)}$ of mode 1 and 2; however, we can use the
results of the
previous section to engineer the state $\varrho_{34}$
to recover the degraded state (see Fig.~\ref{f:scheme}). If $\varrho_{12}$
is initially in a TWB state with TWB parameter $r$, then, by choosing
as $\varrho_{34}$ a TWB with parameter $s=r$ (EntIT
configuration), we have
$\varrho^{(12)} = \varrho_{12}$: the state is totally recovered.
Nevertheless, a partial recover is achieved also when $s<r$ (of course,
if $s>r$ the outgoing state $\varrho^{(12)}$ has properties more
related to $\varrho_{34}$ than to $\varrho_{12}$). This is shown in
Fig.~\ref{f:EfP}, where we plot the $E_f$ and the purity
$\mu = \{(16 \hbox{Det}[{\boldsymbol \Sigma^{(12)}}]\}^{-1/2}$ of
$\varrho^{(12)}$ as functions of $s$ and different values of the
other involved parameters.
\section{Two two-qubit systems invariance via
``remote inversion''}\label{s:RI}
\begin{figure}[tb]
\vspace{0.3cm}
\includegraphics[width=0.35\textwidth]{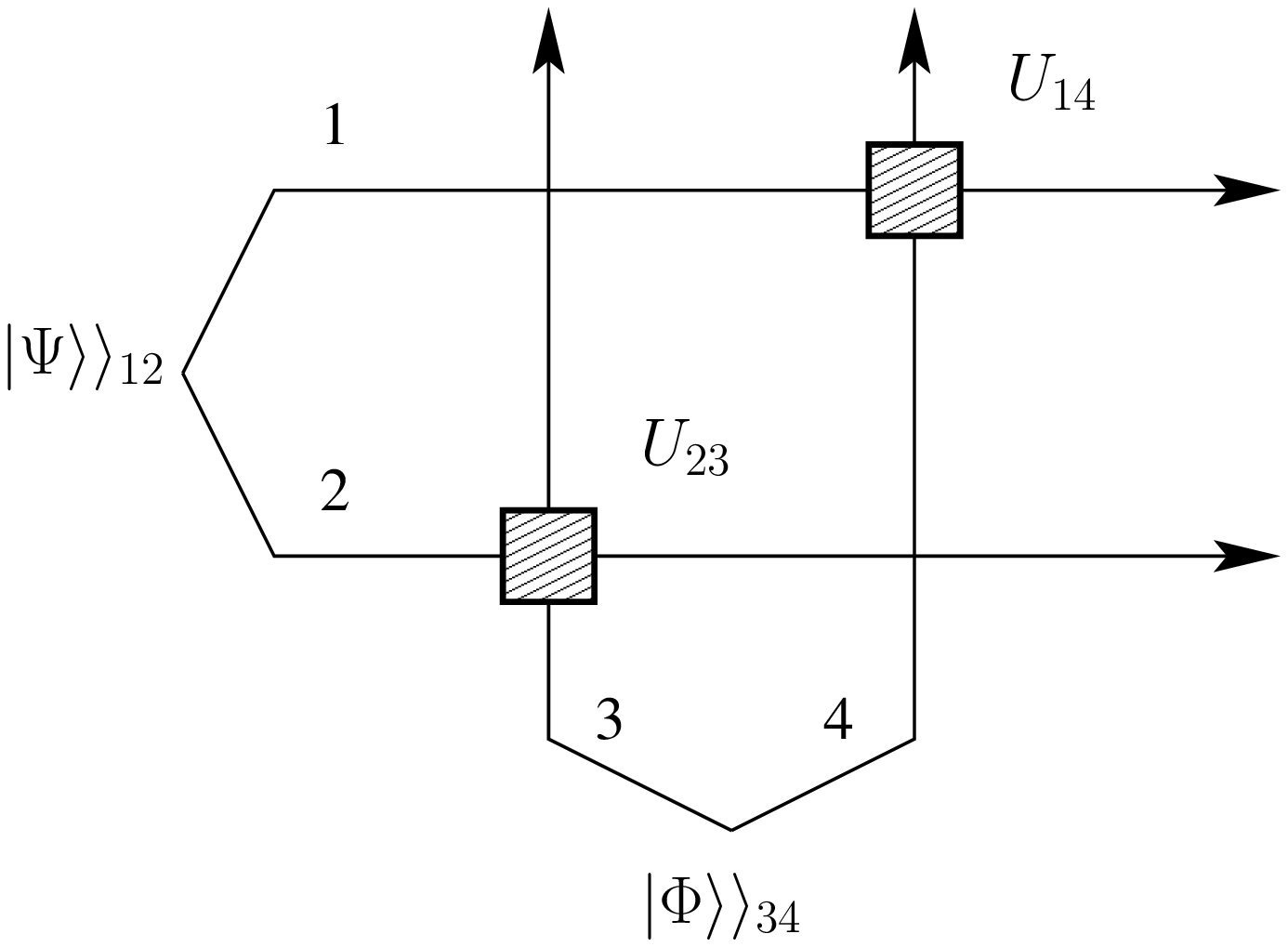}
\vspace{-0.2cm}
\caption{\label{f:QB_scheme} Qubits 14 and 23 of two two-qubit
states undergo the unitary evolutions $U_{14}$ and $U_{23}$, respectively.}
\end{figure}
Let us now consider the qubit couterpart of the continuous variable
setup investigated above.
In this scenario, sketched in Fig.~\ref{f:QB_scheme}, the
qubits 14 and 23 of two two-qubit states $\dket{\Psi}_{12}$ and
$\dket{\Phi}_{34}$ undergo the generic unitary
evolutions $U_{14}$ and $U_{23}$, respectively, where:
\begin{align}
U_{14}({\boldsymbol \theta}) =&
\exp\left( -i \sum_{k=0}^{3}\theta_k\,
\bmsigma_k \otimes \bmsigma_0 \otimes \bmsigma_0 \otimes \bmsigma_k \right), \\
=&\, G_0({\boldsymbol \theta})\,
\bmsigma_0 \otimes \bmsigma_0 \otimes \bmsigma_0 \otimes \bmsigma_0
\nonumber\\
&- \sum_{k=1}^{3} G_k({\boldsymbol \theta})\,
\bmsigma_k \otimes \bmsigma_0 \otimes \bmsigma_0 \otimes \bmsigma_k,\\
U_{23}({\boldsymbol \phi}) =&
\exp\left( -i \sum_{k=0}^{3}\phi_k\,
\bmsigma_0 \otimes \bmsigma_k \otimes \bmsigma_k \otimes \bmsigma_0 \right)\\
 =&\, G_0({\boldsymbol \phi})\,
\bmsigma_0 \otimes \bmsigma_0 \otimes \bmsigma_0 \otimes \bmsigma_0
\nonumber\\
&- \sum_{k=1}^{3} G_k({\boldsymbol \phi})\,
\bmsigma_0 \otimes \bmsigma_k \otimes \bmsigma_k \otimes \bmsigma_0,,
\end{align}
with ${\boldsymbol \theta} = (\theta_0,\theta_1,\theta_2,\theta_3)$,
${\boldsymbol \phi} = (\phi_0,\phi_1,\phi_2,\phi_3)$,
$\bmsigma_k$, $k=0,1,2,3$, are the Pauli matrices, $\bmsigma_0 = {\mathbbm 1}$,
and:
\begin{align}
G_k({\boldsymbol x}) =&\, e^{-i(x_0-x_1-x_2-x_3)}\,\nonumber\\
&\times \left[ 1+ \frac12 \sum_{n,m=1}^3g_k(n,m)\,e^{2i(x_n+x_m)} \right],
\end{align}
${\boldsymbol x}=(x_0,x_1,x_2,x_3)$, with:
\begin{equation}
g_k(n,m) = \left\{
\begin{array}{ll}
+1 & \mbox{if }k\ne n,m \\
0  & \mbox{if }n=m \\
-1 & \mbox{elsewhere}
\end{array}
\right. .
\end{equation}
We are interested in finding the conditions on ${\boldsymbol \theta}$
and ${\boldsymbol \phi}$, that leave the input states unchanged. We
restrict our analysis assuming that $\dket{\Psi}_{12}$ and
$\dket{\Phi}_{34}$ are initially in the same state.
Let us now consider as input states
$\dket{\Psi}_{12} = \frac{1}{\sqrt{2}}\left(
\ket{00}_{12} + \ket{11}_{12} \right)$ and
$\dket{\Phi}_{34} = \frac{1}{\sqrt{2}}\left(
\ket{00}_{34} + \ket{11}_{34} \right)$,
where $\ket{xy}_{kh} = \ket{x}_k\otimes \ket{y}_h$.
The four qubit initial state is then given by:
\begin{align}
| \psi \rangle_{1234} &= \dket{\Psi}_{12} \otimes \dket{\Phi}_{34} \\
&= \frac12\Big(
\ket{00}_{14}\ket{00}_{23} + \ket{01}_{14}\ket{01}_{23} \nonumber\\
&\hspace{0.5cm} + \ket{10}_{14}\ket{10}_{23} + \ket{11}_{14}\ket{11}_{23}\Big)
\label{1234:k}\\
&= \frac12 \Big(
\left|\left.\mbox{$\frac{\bmsigma_0}{\sqrt 2}$}
\right\rangle\!\!\right\rangle_{14}
\left|\left.\mbox{$\frac{\bmsigma_0}{\sqrt 2}$}
\right\rangle\!\!\right\rangle_{23} +
\left|\left.\mbox{$\frac{\bmsigma_1}{\sqrt 2}$}
\right\rangle\!\!\right\rangle_{14}
\left|\left.\mbox{$\frac{\bmsigma_1}{\sqrt 2}$}
\right\rangle\!\!\right\rangle_{23}
\nonumber \\
&\hspace{0.5cm} +
\left|\left.\mbox{$i\frac{\bmsigma_2}{\sqrt 2}$}
\right\rangle\!\!\right\rangle_{14}
\left|\left.\mbox{$i\frac{\bmsigma_2}{\sqrt 2}$}
\right\rangle\!\!\right\rangle_{23} +
\left|\left.\mbox{$\frac{\bmsigma_3}{\sqrt 2}$}
\right\rangle\!\!\right\rangle_{14}
\left|\left.\mbox{$\frac{\bmsigma_3}{\sqrt 2}$}
\right\rangle\!\!\right\rangle_{23}
\Big)
\label{1234:m}
\end{align}
where we re-arranged the tensor product elements in order to put in
evidence the bipartite couples (14 and 23, respectively) involved by
the transformations;
from (\ref{1234:k}) to (\ref{1234:m}), we used the matrix notation
for bipartite states \cite{LoPresti:00}. Now, after some algebra
based on the properties of the Pauli matrices, one can easily find
the following relations:
\begin{subequations}
\label{1234:op}
\begin{align}
&U({\boldsymbol \theta},{\boldsymbol \phi})
\left|\left.\mbox{$\frac{\bmsigma_0}{\sqrt 2}$}
\right\rangle\!\!\right\rangle_{14}
\left|\left.\mbox{$\frac{\bmsigma_0}{\sqrt 2}$}
\right\rangle\!\!\right\rangle_{23} =
\nonumber\\
&\left[G_0({\boldsymbol \theta})-G_1({\boldsymbol \theta})
+G_2({\boldsymbol \theta})-G_3({\boldsymbol \theta})\right]
\\
&\times \left[G_0({\boldsymbol \phi})-G_1({\boldsymbol \phi})
+G_2({\boldsymbol \phi})-G_3({\boldsymbol \phi})\right]
\left|\left.\mbox{$\frac{\bmsigma_0}{\sqrt 2}$}
\right\rangle\!\!\right\rangle_{14}
\left|\left.\mbox{$\frac{\bmsigma_0}{\sqrt 2}$}
\right\rangle\!\!\right\rangle_{23}\nonumber \\
&U({\boldsymbol \theta},{\boldsymbol \phi})
\left|\left.\mbox{$\frac{\bmsigma_1}{\sqrt 2}$}
\right\rangle\!\!\right\rangle_{14}
\left|\left.\mbox{$\frac{\bmsigma_1}{\sqrt 2}$}
\right\rangle\!\!\right\rangle_{23} =
\nonumber\\
&\left[G_0({\boldsymbol \theta})-G_1({\boldsymbol \theta})
-G_2({\boldsymbol \theta})+G_3({\boldsymbol \theta})\right]
\\
&\times\left[G_0({\boldsymbol \phi})-G_1({\boldsymbol \phi})
-G_2({\boldsymbol \phi})+G_3({\boldsymbol \phi})\right]
\left|\left.\mbox{$\frac{\bmsigma_1}{\sqrt 2}$}
\right\rangle\!\!\right\rangle_{14}
\left|\left.\mbox{$\frac{\bmsigma_1}{\sqrt 2}$}
\right\rangle\!\!\right\rangle_{23}\nonumber \\
&U({\boldsymbol \theta},{\boldsymbol \phi})
\left|\left.\mbox{$i\frac{\bmsigma_2}{\sqrt 2}$}
\right\rangle\!\!\right\rangle_{14}
\left|\left.\mbox{$i\frac{\bmsigma_0}{\sqrt 2}$}
\right\rangle\!\!\right\rangle_{23} =
\nonumber\\
&\left[G_0({\boldsymbol \theta})+G_1({\boldsymbol \theta})
+G_2({\boldsymbol \theta})+G_3({\boldsymbol \theta})\right]
\\
&\times\left[G_0({\boldsymbol \phi})+G_1({\boldsymbol \phi})
+G_2({\boldsymbol \phi})+G_3({\boldsymbol \phi})\right]
\left|\left.\mbox{$i\frac{\bmsigma_2}{\sqrt 2}$}
\right\rangle\!\!\right\rangle_{14}
\left|\left.\mbox{$i\frac{\bmsigma_2}{\sqrt 2}$}
\right\rangle\!\!\right\rangle_{23}\nonumber \\
&U({\boldsymbol \theta},{\boldsymbol \phi})
\left|\left.\mbox{$\frac{\bmsigma_3}{\sqrt 2}$}
\right\rangle\!\!\right\rangle_{14}
\left|\left.\mbox{$\frac{\bmsigma_3}{\sqrt 2}$}
\right\rangle\!\!\right\rangle_{23} =
\nonumber\\
&\left[G_0({\boldsymbol \theta})+G_1({\boldsymbol \theta})
-G_2({\boldsymbol \theta})-G_3({\boldsymbol \theta})\right]
\\
&\times\left[G_0({\boldsymbol \phi})+G_1({\boldsymbol \phi})
-G_2({\boldsymbol \phi})-G_3({\boldsymbol \phi})\right]
\left|\left.\mbox{$\frac{\bmsigma_3}{\sqrt 2}$}
\right\rangle\!\!\right\rangle_{14}
\left|\left.\mbox{$\frac{\bmsigma_3}{\sqrt 2}$}
\right\rangle\!\!\right\rangle_{23}\nonumber
\end{align}
\end{subequations}
where we defined
$U({\boldsymbol \theta},{\boldsymbol \phi}) \equiv
U_{14}({\boldsymbol \theta}) U_{23}({\boldsymbol \phi})$
and used the property $A\otimes B\dket{C} = \dket{ACB^T}$ of the matrix
representation of bipartite states. Hence, the input states
are left unchanged if the condition
${\boldsymbol \phi} = -{\boldsymbol \theta}$ is met: in this case all the
numerical coefficients appearing at the right hand sides of
Eq.s~(\ref{1234:op}) are equal to 1. The same result holds if we
consider as input states $\dket{\Psi}_{12}$ and $\dket{\Phi}_{34}$
one of the three left Bell states.
For a generic two qubit state the previous condition is not enough to
leave it unchanged, and more condition on ${\boldsymbol \theta}$ are
required (see Appendix~\ref{A:zoology} for a complete zoology).
\par
We can also look at the invariance obtained for Bell states as follows.
The action of an unitary transformation, acting on two qubits belonging to
different couples of qubits initially in the same Bell state, can be
canceled out by applying the inverse transformation to the remaining couple
of qubits (choosing ${\boldsymbol \phi} = -{\boldsymbol \theta}$ formally
corresponds to the inverse of the first transformation, of course, acting
on a different system). For this effect (invariance), Bell states plays a
crucial role: if we consider as starting states other than Bell states,
the inversion of the operation is not enough to achieve the invariance,
further conditions are needed, i.e., differently form the Bell state
case, not all the classes of unitaries leads to invariance up to
``remote inversion''.
\section{Conclusions}\label{s:remarks}
In this paper we have investigated in some details a useful symmetry 
exhibited by pairs of entangled states, which induces operation 
transparency, {\em i.e} the preservation of  the state under the 
action of specific class of unitaries.
\par
In continuous variable systems 
entanglement induced transparency occurs when two TWBs with the same 
energy are left unchanged after the evolution through two equal 
beam splitters. We have shown that entanglement is crucial
for the effect and we have studied its occurrence and robustness.
Besides, we have shown how EntIT may be useful to engineer baths with 
nonclassical correlations in order to preserve transmission of entanglement 
and the purity of TWBs during the propagation.
Related to the EntIT effect is the double swapping: now
entanglement is swapped between the modes of TWBs by simply changing
the phase of the initial bipartite states before the interaction at
two balanced BSs and without any measurement. 
\par
The investigation of the discrete-variable counterpart of the EntIt
has brought us to the ``remote inversion'' effect, i.e., the action of
an unitary transformation, acting on two qubits belonging to different
couples of qubits initially in the same Bell state, can be canceled
out by applying the inverse transformation to the remaining couple of
qubits. This effect my be used to remotely control quantum operations
over a quantum network.
\acknowledgments
Discussions with M.~Genoni, P.~Giorda, S.~Mancini and A.~R.~Rossi
are acknowledged. This work has been partially supported by the CNR-CNISM
convention.
\appendix
\section{Zoology for qubit invariance via
``remote inversion''}\label{A:zoology}
Here we assume that both the states $\dket{\Psi}_{12}$ and
$\dket{\Phi}_{34}$ are equal to the same state:
\begin{equation}
\dket{\psi} = \sum_{h,k=0,1} a_{hk} \ket{hk},
\end{equation}
with $\sum_{h,k} a_{hk}^2 = 1$ (without loss of generality we choose
only real coefficients).
We have seen in Section \ref{s:RI} that if $\dket{\psi}$ is one of the
four Bell states, then the invariance is achieved for
${\boldsymbol \phi} = -{\boldsymbol \theta}$. In all the other
cases the following further conditions are needed (of course, together
with ${\boldsymbol \phi} = -{\boldsymbol \theta}$).
\begin{itemize}
\item $a_{00}=1$ or $a_{11}=1$ $\Rightarrow$ $\theta_1=\theta_2$.
\item $a_{01}=1$ or $a_{10}=1$ $\Rightarrow$ $\theta_1=-\theta_2$.
\item $a_{00},a_{11}\ne 0$, $a_{00}\ne a_{11}$, and $a_{01},a_{10}= 0$
  $\Rightarrow$ $\theta_1=\theta_2$.
\item $a_{01},a_{10}\ne 0$, $a_{01}\ne a_{10}$, and $a_{00},a_{11}= 0$
  $\Rightarrow$ $\theta_1=-\theta_2$.
\item $a_{hk} \ne 0, \forall h,k$:
\begin{itemize}
\item  $a_{00} = a_{11}$ and  $a_{01} = a_{10}$
  $\Rightarrow$ $\theta_2=\theta_3$.
\item  $a_{00} = -a_{11}$ and  $a_{01} = -a_{10}$
  $\Rightarrow$ $\theta_2=-\theta_3$.
\item  $a_{00} = -a_{11}$ and  $a_{01} = a_{10}$
  $\Rightarrow$ $\theta_1=\theta_3$.
\item  $a_{00} = a_{11}$ and  $a_{01} = -a_{10}$
  $\Rightarrow$ $\theta_1=-\theta_3$.
\end{itemize}
\item In all the other cases one should put
  ${\boldsymbol \phi} = {\boldsymbol \theta} = 0$,
  i.e., the states are left unchanged only if no operation is performed.
\end{itemize}


\begin{thebibliography}{30}
\bibitem{eres} W. K. Wootters, Phil. Trans. R. Soc. Lond. A {\bf 356},
1717 (1998).
\bibitem{edet} O. Guhne, G. Toth, Physics Reports {\bf 474}, 1 (2009).
\bibitem{eman}J. M. Raimond, M. Brune, S. Haroche, Rev. Mod. Phys. {\bf
73}, 565 (2001).
\bibitem{BB1} D.~Vitali and P.~Tombesi, Phys Rev A {\bf 59}, 4178 (1999).
\bibitem{BB2} S.~Damodarakurup, et a., 	arXiv:0811.2654v1 [quant-ph].
\bibitem{DFS97} P.~Zanardi and M.~Rasetti, Mod. Phys. Lett. B {\bf 11}, 1085 (1997);
P. Zanardi and M. Rasetti, Phys. Rev. Lett. 79, 3306 (1997).
\bibitem{cl1} H.~J.~Briegel and R.~Raussendorf, Phys. Rev. Lett. {\bf 86},
910 (2001).
\bibitem{aga86} G.~S.~Agarwal, Phys. Rev. Lett. {\bf 57}, 827 (1986).
\bibitem{aga05} G.~S.~Agarwal  and A.~Biswas, J. Opt B {\bf 7}, 350 (2005).
\bibitem{us1} V. C. Usenko, B. I. Lev, Phys. Lett. A {\bf 348}, 17
(2005).
\bibitem{us2} V. C. Usenko, M. G. A. Paris, Phys. Rev. A {\bf 75},
043812 (2007).
\bibitem{gsb} J. Eisert \emph{et al.}, Int. J. Quant. Inf.  \textbf{1}, 479 (2003); A.~Ferraro 
\emph{et al.}, {\em Gaussian States in Quantum Information}, (Bibliopolis,
Napoli, 2005); G. Adesso \emph{et al.}, J. Phys. A 40, 7821 (2007); 
S.~L.~Braunstein \emph{et al.}, Rev. Mod. Phys. \textbf{77},513 (2005); 
\bibitem{EOFM} P. Marian and T. A. Marian, Phys. Rev. Lett. {\bf 101}, 220403
(2008).
\bibitem{CM02} C.~Macchiavello and G.~M.~Palma, Phys. Rev. A {\bf 65},
050301(R) (2002).
\bibitem{KB04} K. Banaszek, A. Dragan, W. Wasilewski and C. Radzewicz, Phys. Rev. Lett. {\bf 92}, 257901 (2004).
\bibitem{DK05} D.~Kretschmann and R.~F.~Werner, Phys. Rev. A {\bf 72},
062323 (2005).
\bibitem{NJC05} N.~J.~Cerf, J. Clavareau, C.Macchiavello and J. Roland, Phys. Rev. A {\bf 72}, 042330 (2005).
\bibitem{CL09} C.~Lupo and S.~Mancini, arXiv:0901.4966v1 [quant-ph].
\bibitem{TABK88} T.~A.~B.~Kennedy and D.~F.~Walls, Phys. Rev. A {\bf 37}, 152 (1988).
\bibitem{PT94} P.~Tombesi and D.~Vitali,  Phys. Rev. A {\bf 50}, 4253 (1994).
\bibitem{NL98} N.~L\"{u}tkenhaus, J.~I.~Cirac and P.~Zoller, Phys. Rev. A {\bf 57},
548 (1998).
\bibitem{SO03} S.~Olivares, M.~G.~A.~Paris and A.~R.~Rossi, Phys. Lett. A
{\bf 319}, 32 (2003); A.~R.~Rossi, S.~Olivares and M.~G.~A.~Paris,
J. Mod. Opt. {\bf 51}, 1057 (2004).
\bibitem{LoPresti:00} G.~M.~D'Ariano, P.~Lo Presti, and M.~F.~Sacchi,
Phys. Lett. A {\bf 272}, 23 (2000).
\end{thebibliography}
\end{document}